\documentclass[12pt]{article}
\usepackage{citesort}
\usepackage{graphicx}
\usepackage{amssymb}

\newcommand{\nc}{\newcommand}
\nc{\renc}{\renewcommand}

\nc{\half}{{\textstyle{1\over2}}}
\nc{\etal}{\mbox{\it et al. }}
\nc{\ie}{{\it i.e.}}
\nc{\eg}{{\it e.g.}}

\renc{\thefootnote}{\arabic{footnote}}
\nc{\capt}[1]{{\bf Figure.} {\small\sl #1}}
\nc{\eqs}[2]{\mbox{Eqs.~(\ref{#1},\,\ref{#2})}}
\nc{\eq}[1]{\mbox{Eq.~(\ref{#1})}}

\nc{\figs}[2]{\mbox{Figs.~(\ref{#1},\,\ref{#2})}}
\nc{\fig}[1]{\mbox{Fig~.(\ref{#1})}}

\nc{\eea}{\vspace{\undereqskip}\end{eqnarray}}
\nc{\ee}{\vspace{\undereqskip}\end{equation}}
\nc{\bdm}{\begin{displaymath}}
\nc{\edm}{\end{displaymath}}
\nc{\dpsty}{\displaystyle}
\nc{\bc}{\begin{center}}
\nc{\ec}{\end{center}}
\nc{\ba}{\begin{array}}
\nc{\ea}{\end{array}}
\nc{\bab}{\begin{abstract}}
\nc{\eab}{\end{abstract}}
\nc{\btab}{\begin{tabular}}
\nc{\etab}{\end{tabular}}
\nc{\bit}{\begin{itemize}}
\nc{\eit}{\end{itemize}}
\nc{\ben}{\begin{enumerate}}
\nc{\een}{\end{enumerate}}
\nc{\bfig}{\begin{figure}}
\nc{\efig}{\end{figure}}
\nc{\arreq}{&\!=\!&}
\nc{\arrmi}{&\!-\!&}
\nc{\arrpl}{&\!+\!&}
\nc{\arrap}{&\!\!\!\approx\!\!\!&}
\nc{\non}{\nonumber\\*}
\nc{\align}{\!\!\!\!\!\!\!\!&&}
\def\lsim{\; \raise0.3ex\hbox{$<$\kern-0.75em
      \raise-1.1ex\hbox{$\sim$}}\; }
\def\gsim{\; \raise0.3ex\hbox{$>$\kern-0.75em
      \raise-1.1ex\hbox{$\sim$}}\; }
\nc{\DOT}{\hspace{-0.08in}{\bf .}\hspace{0.1in}}
\nc{\Laada}{\hbox {$\sqcap$ \kern -1em $\sqcup$}}
\nc\loota{{\scriptstyle\sqcap\kern-0.55em\hbox{$\scriptstyle\sqcup$}}}
\nc\Loota{{\sqcap\kern-0.65em\hbox{$\sqcup$}}}
\nc\laada{\Loota}
\nc{\qed}{\hskip 3em \hbox{\BOX} \vskip 2ex}

\nc{\real}{{\rm I \! R}}
\nc{\Z}{{\sf Z \!\!\! Z}}
\nc{\complex}{{\rm C\!\!\! {\sf I}\,\,}}
\def\bigid{\leavevmode\hbox{\small1\kern-3.8pt\normalsize1}}
\def\id{\leavevmode\hbox{\small1\kern-3.3pt\normalsize1}}
\nc{\slask}{\!\!\!/}
\nc{\bis}{{\prime\prime}}
\nc{\pa}{\partial}
\nc{\na}{\nabla}
\nc{\ra}{\rangle}
\nc{\la}{\langle}
\nc{\goto}{\rightarrow}
\nc{\swap}{\leftrightarrow}

\nc{\EE}[1]{ \mbox{$\cdot10^{#1}$} }
\nc{\abs}[1]{\left|#1\right|}
\nc{\at}[2]{\left.#1\right|_{#2}}
\nc{\norm}[1]{\|#1\|}
\nc{\abscut}[2]{\Abs{#1}_{\scriptscriptstyle#2}}
\nc{\vek}[1]{{\rm\bf #1}}
\nc{\integral}[2]{\int\limits_{#1}^{#2}}
\nc{\inv}[1]{\frac{1}{#1}}
\nc{\dd}[2]{{{\partial #1}\over{\partial #2}}}
\nc{\ddd}[2]{{{{\partial}^2 #1}\over{\partial {#2}^2}}}
\nc{\dddd}[3]{{{{\partial}^2 #1}\over
        {\partial #2 \partial #3}}}
\nc{\dder}[2]{{{d #1}\over{d #2}}}
\nc{\ddder}[2]{{{d^2 #1}\over{d {#2}^2}}}
\nc{\dddder}[3]{{d^2 #1}\over
        {d #2 d #3}}
\nc{\dx}[1]{d\,^{#1}x}
\nc{\dy}[1]{d\,^{#1}y}
\nc{\dz}[1]{d\,^{#1}z}
\nc{\dl}[1]{\frac{d\,^{#1}l}{(2\pi)^{#1}}}
\nc{\dk}[1]{\frac{d\,^{#1}k}{(2\pi)^{#1}}}
\nc{\dq}[1]{\frac{d\,^{#1}q}{(2\pi)^{#1}}}

\nc{\cc}{\mbox{$c.c.$ }}
\nc{\hc}{\mbox{$h.c.$ }}
\nc{\cf}{cf.\ }
\nc{\erfc}{{\rm erfc}}
\nc{\Tr}{{\rm Tr\,}}
\nc{\tr}{{\rm tr\,}}
\nc{\pol}{{\rm pol}}
\nc{\sign}{{\rm sign}}
\nc{\bfT}{{\bf T }}

\nc{\cA}{{\cal A}}
\nc{\cB}{{\cal B}}
\nc{\cD}{{\cal D}}
\nc{\cE}{{\cal E}}
\nc{\cG}{{\cal G}}
\nc{\cH}{{\cal H}}
\nc{\cL}{{\cal L}}
\nc{\cO}{{\cal O}}
\nc{\cT}{{\cal T}}
\nc{\cN}{{\cal N}}
\nc{\rvac}[1]{|{\cal O}#1\rangle}
\nc{\lvac}[1]{\langle{\cal O}#1|}
\nc{\rvacb}[1]{|{\cal O}_\beta #1\rangle}
\nc{\lvacb}[1]{\langle{\cal O}_\beta #1 |}
\nc{\bb}{\bar{\beta}}
\nc{\bt}{\tilde{\beta}}
\nc{\ctH}{\tilde{\cal H}}
\nc{\chH}{\hat{\cal H}}
\nc{\1}{\aa}
\nc{\2}{\"{a}}
\nc{\3}{\"{o}}
\nc{\4}{\AA}
\nc{\5}{\"{A}}
\nc{\6}{\"{O}}
\nc{\al}{\alpha}
\nc{\g}{\gamma}
\nc{\Del}{\Delta}
\nc{\e}{\epsilon}
\nc{\eps}{\epsilon}
\nc{\lam}{\lambda}
\nc{\om}{\omega}
\nc{\Om}{\Omega}
\nc{\ve}{\varepsilon}
\nc{\mn}{{\mu\nu}}
\nc{\ka}{\kappa}
\nc{\vp}{\varphi}

\begin{document}
{\title{\vskip-2truecm{\hfill {{\small HIP-2004-26/TH\\ \hfill \\
        }}\vskip 1truecm} {\LARGE Inflation in large $N$ limit of 
supersymmetric gauge theories}}}
\vspace{-.2cm}
{\author{
{\sc Asko Jokinen$^1$ and Anupam Mazumdar$^2$}
\\
{\sl $^1$\small Helsinki Institute of Physics, P.O. Box 64, FIN-00014
        University of Helsinki, FINLAND}
\\
{\sl $^2$\small McGill University, 3600 University Road, Montr\'eal, Qu\'ebec,
        H3A 2T8, CANADA}}}
\maketitle

\begin{abstract}
Within supersymmetry we provide an example where the inflaton sector
is derived from a gauge invariant polynomial of $SU(N)$ or $SO(N)$ gauge
theory.  Inflation in our model is driven by multi-flat directions,
which assist accelerated expansion. We show that multi-flat directions
can flatten the individual non-renormalizable potentials such that
inflation can occur at sub-Planckian scales. We calculate the density
perturbations and the spectral index, we find that the spectral index
is closer to scale invariance for large $N$. In order to realize a successful
cosmology we require large $N$ of order, $N\sim 600$.
\end{abstract}

%%%%%%%%%%%%%%%%%%%%%%%%%%%%%%%%%
\section{Introduction}

The current satellite based experiments suggest that the early
universe might have had a spell of accelerated expansion~\cite{WMAP}.
The idea of inflation is interesting as it solves quite a range of
problems~\cite{Guth}. Usually in the literature it is assumed that a
single gauge singlet with its non-vanishing potential energy is
responsible for driving inflation, see~\cite{Lyth}. However a gauge
singlet is a step backward in presenting cosmological inflationary
models, because the gauge couplings and the masses can be tweaked at
our will, or from the observational constraints, and not from any
fundamental gauged sector~\footnote{In some models of inflation the
Higgs potential is responsible for providing the non-zero vacuum
expectation value during inflation. However inflation in these models
can last forever if there is no first or second order phase
transition. The phase transition leads to an end of inflation due to a
rolling scalar field which is usually treated as a gauge singlet
inflaton~\cite{Linde}. This particular feature is explicit in a
supersymmetric set up where the total potential, within
$N=1$~supersymmetry (SUSY), is obtained from the superpotential,
$W=\lambda\Phi (H^2-H_{0})^2$, where $\Phi$ is the absolute gauge
singlet and $H$ is the Higgs sector.}. The lack of motivation behind
the absolute gauge singlet in nature enforces gauging the inflaton
sector. It is equally appealing if we wish to connect the inflaton
sector to a realistic particle physics model.

Supersymmetric gauge theories bring a new host of flat directions
along which the scalar potential vanishes. Within $N=1$ SUSY it
implies that D-term and F-term vanishes identically. These flat
directions are derived from gauge invariant polynomials
\cite{Buccella:1981ib,Dine:1995kz,Gherghetta:1995dv}, which are build
on gauge invariant combination of squarks and sleptons. Interestingly
SUSY flat directions do not receive any perturbative corrections
\cite{Seiberg:1993vc}. However they can obtain non-perturbative
superpotential and K\"ahler corrections when supersymmetry is
broken~\cite{gwr79,Dine:1995kz}. These flat directions have many
cosmological implications, such as baryogenesis~\cite{Dine:1995kz},
dark matter~\cite{Kusenko}, and as a potential source for the cosmic density
perturbations~\cite{Enqvist}, for a review on the subject,
see~\cite{Enqvist:2003gh}. In this paper our main goal is to seek a
gauge invariant flat direction(s) as a candidate for the inflaton
within minimal supersymmetric standard model (MSSM) and/or beyond MSSM.

We will explain the caveats in treating MSSM flat directions as an
inflaton. MSSM flat directions are usually parameterized by a
monomial, which does not lead to any inflation because
non-renormalizable potentials dominate at large vacuum expectation
values (vevs), whose contributions can only be trusted below their
cut-off. It is helpful to represent the flat direction by gauge
invariant polynomials. This brings a new host of flat directions,
see~\cite{Enqvist:2003pb}. We will argue that spanning the moduli
space of flat directions can lead to inflation due to a collective
motion of these flat directions in a moduli space.

The concept of having inflation from multi-fields is well known.  In
Ref.~\cite{Liddle:1998jc}, it was first demonstrated that inflation
occurs at ease with many exponentials, and it has been generalized to
many other forms of potentials, see~\cite{Assist}.

In this paper we will argue that enhancing the gauge group along with
the matter content will provide multi-flat directions within $SU(N)$
and/or $SO(N)$ gauge theories, where inflation will be driven collectively by
the degrees of freedom below the cut-off scale, which we consider here as the
Planck scale.

%%%%%%%%%%%%%%%%%%%%%%%%%%%%%%%%%%%%%%%%%%%%%%%%%%%%%%%%%%%%%%%%%%%%%%%
\section{Why is it hard to obtain inflation from MSSM flat directions?}

Within MSSM there are many flat directions subject to F and D
constraints
\begin{equation}
F_{i}\equiv \frac{\partial W}{\partial\Phi_{i}}=0\,,~~D^{A}\equiv 
\Phi^{\dagger}T^{A}\Phi=0\,,
\end{equation}
for the scalars $\Phi_{i}$. More elegant way of describing a flat
direction is through gauge invariant holomorphic polynomials of the
chiral superfields $\Phi_{i}$. Within MSSM, with R-parity, all the
flat directions have been tabulated, see~\cite{Gherghetta:1995dv}. In
a cosmological context where supersymmetry is broken by the finite
energy density of the Universe, the flat directions obtain various
corrections. Notably the non-renormalizable superpotential corrections
are of types~\cite{Seiberg:1993vc,Dine:1995kz}
\begin{equation}
W\sim \lambda \frac{\Phi^{n}}{M_{p}^{n-3}}\,,~~~~
W\sim \lambda_1\frac{\Phi^{n-1}\Psi}{M_{p}^{n-3}}\,,
\end{equation}
where, $\Phi,~\Psi$ are flat directions, $\lambda,\lambda_1\sim {\cal
O}(1)$, and $M_{p}\sim 10^{18}$~GeV.  Within MSSM, with R-parity,
all the flat directions are lifted by the non-renormalizable
operators, $n=4,5,6,7,9$, see~\cite{Gherghetta:1995dv}.

Besides such a non-renormalizable correction, the MSSM flat
directions naturally obtain soft SUSY breaking mass terms,
\begin{equation}
\label{chaotic1}
V\sim m_{soft}^2\Phi^2\,,
\end{equation}
where $m_{soft}\sim {\cal O}(1)$~TeV. Note that the above potential is
similar to that of a chaotic type potential~\cite{Linde1}, but with
a mass parameter which is way too small to provide any observable
effects, since $\delta\rho/\rho\sim m_{soft}/M_{p}\sim {\cal O}({\rm
TeV}/M_{p})\sim 10^{-15}$. Besides COBE/WMAP normalization requires that
$m_{soft}\sim 10^{13}$ GeV~\cite{Lyth:1998xn}, which is ten orders of 
magnitude larger than what we expect from soft SUSY breaking mass term.

At sufficiently large vevs, $\Phi,\Psi\gg {\cal O}(\rm TeV)$, the
potential from the non-renor\-ma\-li\-zab\-le superpotential terms dominate
over the soft SUSY breaking potential. One might expect to realize
inflation from the MSSM flat directions with a potential derived 
from Eq.~(\ref{chaotic1}),
\begin{equation}
\label{chaotic2}
V\sim |\lambda|^2\frac{\Phi^{2(n-1)}}{M_{p}^{2n-6}}\,.
\end{equation}
The above potential mimics that of a chaotic type inflationary
model~\cite{Linde1}. Inflation with such a potential is possible only
when the vev is larger than the cut-off. Inflation ends when
$\Phi_{end}\sim (2n-2)M_{p}$. However in our case, since $\Phi$ is a
gauge invariant quantity, we can no longer trust the
non-renormalizable potential above the cut-off with a coefficient,
$\lambda\sim {\cal O}(1)$. Further note that even if we take the vev
larger than $M_{p}$, the potential is too steep to give rise an
interesting effect. Furthermore the coupling constant should be
$\lambda \lsim 10^{-14}$~\cite{Lyth:1998xn}, in order to provide the
right amplitude for the density perturbations, which is ridiculously
small compared to any SM/MSSM Yukawa and/or gauge couplings.
These facts immediately suggest that inflation with the MSSM
flat directions is very unlikely. The main challenge is to realize
inflation at vevs smaller than the Planck scale, which is impossible
to realize with a single flat direction.

Besides the superpotential contribution, the flat direction also
obtains correction due to the K\"ahler potential in $N=1$
supergravity. For a minimal choice of K\"ahler potential,
$K=\pm\Phi_{i}^{\dagger}\Phi_{i}+...$, the flat direction obtains a
potential,
\begin{equation}
V\sim \pm {\cal O}(1)H^2\Phi^2\,,
\end{equation}
where $H$ is the Hubble rate during and after inflation. However there
is a bit of freedom in the choice of a K\"ahler potential. Usually
such corrections are not present in No-Scale type k\"ahler
models~\cite{Lahanas:1986uc}. For our purpose we will ignore such a
dangerous correction which leads to an unsuccessful inflationary
scenario. This is because a positive Hubble induced mass correction
will spoil any interesting dynamics.

Motivated by these problems we are inclined to study the properties of
multi flat directions, which will evolve collectively in such a way to
assist inflation similar to the assisted inflationary
scenarios~\cite{Liddle:1998jc}~\footnote{If many scalar fields evolve
independently then they assist inflation inspite of the fact that
individual potential is unable to sustain inflation on its own. The
key point is that collective dynamics of fields increase the Hubble
friction term which leads to the slow rolling of the fields. The main
constraint is that the fields ought to be free, the coupling between
the fields do not assist inflation~\cite{Liddle:1998jc,Assist}. }. One of the
revealing properties of the assisted inflation is that it is possible
to drive inflation for chaotic type potentials, $V\sim
\sum_{i}\phi_{i}^{n}$, with vevs below the Planckian scale, see 
also~\cite{Kanti}. This is a good news for us, because the
non-renormalizable potentials are trusted only below the
cut-off. Therefore there is a glimmer of hope that we might be able to
sustain inflation from the flat directions with potentials of type
Eq.~(\ref{chaotic2}), albeit with many directions.

However the question is to seek whether MSSM can provide us with
sufficient number of independent flat directions? For example,
$LH_{u},~udd$, directions are lifted by $n=4$ non-renormalizable
operators, which can obtain large vevs simultaneously, but
$LH_{u},~LLe$ directions, depending on the family indices, need not be
simultaneously flat.  As we shall show in section $IV$ we would
require at least $600$ independent directions. Unfortunately within
MSSM there are only $334$ D-flat directions. This number is further
reduced due to the F-term constraints, and moreover these directions
are not independent to each other.

So much for a flat direction represented by monomials. Very recently
we explored, for the first time, the dynamics of multiple flat
directions parameterized by the gauge invariant polynomials in
Ref.~\cite{Enqvist:2003pb}. It turns out that the multiple flat
directions have very interesting properties. We mention them briefly;
for a single flat direction the choice of a gauge is trivially
satisfied to be a pure gauge by definition, see for detailed
discussion in Ref.~\cite{Enqvist:2003pb}. However the choice of a pure
gauge in not trivially satisfied in the multi-flat direction case,
unless one assumes that the gauge degrees of freedom are frozen with a
homogeneous distribution without any spatial perturbations. Any
spatial fluctuation will excite the gauge degrees of freedom with an
interesting astrophysical implication, such as exciting the
seed magnetic field, see~\cite{ejmmag}.

In Ref.~\cite{Enqvist:2003pb}, we studied a polynomial $I$ spanned by
the Higgses and the sleptons,
\begin{equation}
\label{polyn1}
I = \nu_1 H_u L_1 + \nu_2 H_u L_2 + \nu_3 H_u L_3
\end{equation}
where $\nu_i$ are complex coefficients, $H_u$ is the up-type Higgs and
$L_i$ are the sleptons. For a following field configuration, the polynomial
$I$ has a vanishing matter current and vanishing gauge fields
\cite{Enqvist:2003pb},
\begin{equation}
\label{struct321}
L_i = e^{-i\chi/2} \phi_i \left( \ba{ll} 0 \\ 1 \ea \right), \quad H_u =
e^{i\chi/2} \sqrt{\sum_i |\phi_i|^2} \left( \ba{ll} 1 \\ 0 \ea \right)\,,
\end{equation}
where $\phi_i$ are complex scalar fields and the phase $\chi$ is a
real field constrained by
\begin{equation}
\label{constr320}
\partial_{\mu} \chi = \frac{\sum_j J_j^\phi}{2i \sum_k |\phi_k|^2}\,, \qquad
J_i^{\phi} = \phi_i^* \partial_{\mu} \phi_i - \phi_i \partial_{\mu} \phi_i^*
\,.
\end{equation}
The field configuration in Eq.~(\ref{struct321}) leads to an effective
Lagrangian for the flat direction fields $\phi_i$,
\begin{equation}
\label{lagr324}
\mathcal{L} = |D_{\mu} H_u|^2 + \sum_{i=1}^3|D_{\mu} L_i|^2 -V =
\frac{1}{2} \partial_{\mu} \Phi^{\dagger} \left(1 + P_1 - \frac{1}{2}P_2
\right) \partial^{\mu} \Phi - V\,,
\end{equation}
where $D_{\mu}$ is a gauge covariant derivative that reduces to the
partial derivative when the gauge fields vanish, $P_1$ is the
projection operator along $\Phi$ and $P_2$ along $\Psi$, where
\begin{eqnarray}
\bar{\phi} = \left( \phi_1 \quad \phi_2 \quad \phi_3 \right)^T \,, \qquad 
\Phi = \left( \begin{array}{ll} \bar{\phi} \\ \bar{\phi}^*
  \end{array} \right) \,, \qquad \Psi = \left( \begin{array}{ll}
    \,\,\bar{\phi} \\ -\bar{\phi}^* \end{array} \right) \,,
\end{eqnarray}
and the corresponding equation of motion
\begin{eqnarray}
\label{eqm327}
\partial_{\mu}\partial^{\mu}\Phi + 3H\dot\Phi +
\left(1-\frac{1}{2}P_1+P_2\right) \dd{V}{\Phi^{\dagger}}
- R^{-2} \left[\partial_{\mu}\Psi\,(\Psi^{\dagger}
  \partial^{\mu}\Phi)\right. & & \nonumber \\ + \left.
  \Psi\,(\partial_{\mu}\Psi^{\dagger} P_2 \partial^{\mu}\Phi) +
  \frac{1}{2}\Phi\,\partial_{\mu} \Phi^{\dagger}
  \left(1-P_1-\frac{3}{2}P_2\right) \partial^{\mu}\Phi \right] &=& 0,
\end{eqnarray}
where $R=\sqrt{\Phi^{\dagger}\Phi}$. We are interested in the
background dynamics where all the fields are homogeneous in time, and
for simplicity we study only the radial motion, such that $\Phi = R
\hat{e}_{\Phi}$, where $\dot{\hat{e}}_{\Phi}=0$ (the dot denotes 
derivative w.r.t time). Then the equation of motion simplifies to
\begin{equation}
\ddot R + 3H\dot R + \frac{1}{2} \frac{\partial V}{\partial R} = 0\,,
\end{equation}
if we assume $V=V(R)$ for simplicity. A notable feature is that the
fields have non-minimal kinetic terms, since the field manifold
defined by the flat direction is curved, actually a hyperbolic
manifold. This results into the usual equation of motion for one
scalar field with a potential for the radial mode except for the
factor $1/2$, which makes the potential effectively flatter in this
direction. This can be traced to the square root nature of $H_u$ in
Eq.~(\ref{struct321}).

This illustrates that there is a way of making the flat direction
potential even flatter in the equation of motion. Now one might, of
course, wonder whether MSSM flat directions might work. As far as our
example of $LH_{u}$ is concerned there are only three families which
we can account for. The flattest MSSM direction, $QuQue$, is lifted by
$n=9$ superpotential operator, $QuQuQuH_{d}ee$. The flat direction
$QuQue$ is an 18 complex dimensional manifold as can be seen from
Table 5 of Ref.~\cite{Gherghetta:1995dv}. The largest D-flat direction
is only 33 complex dimensional \cite{Gherghetta:1995dv}. As we shall
see in section $IV$, we would require much larger number of fields to
assist inflation below the Planck vev. The conclusion is that we must
go beyond the MSSM gauge group to seek a large number of flat
directions.

%%%%%%%%%%%%%%%%%%%%%%%%%%%%%%%%%%%%%%%%%%%%%%%%%%%%%%%%%%%%%%%%%%%%%%%
\section{Construction of flat directions in $SU(N)$ and \\ $SO(N)$ gauge
  theories}

Now we wish to study the large $N$ gauge group which could provide us
with multi flat directions. We do not pretend here to scan all the D-
and F- flat directions of either $SU(N)$ or $SO(N)$, but we take a
particular combination which is indeed a flat direction. Encouraged by
our previous study within MSSM, we consider $M$ fields $H_i$ and $N-1$
fields $G_j$ in the fundamental representation $N$ of the gauge
group. Note that the matter content is also enhanced, which has a total
$N-1+M$ degrees of freedom.  Then there exists a D-flat direction
described by a gauge invariant polynomial
\begin{equation}
\label{polyn}
I=\sum_{j=1}^M \alpha_j \eps_{d_1\cdots d_{N-1} e} H_1^{d_1}\cdots
H_{N-1}^{d_{N-1}} G_j^e\,,
\end{equation}
which after solving the constraint equations
\begin{equation}
\label{flat}
\frac{\partial I}{\partial H_j^a} = C H_j^{a*}, \qquad \frac{\partial
  I}{\partial G_i^a} = C G_i^{a*}
\end{equation}
produces a vacuum configuration
\begin{eqnarray}
\label{vacuum}
H_j^a &=& \delta_N^a \phi_j \qquad \qquad \qquad j=1,\ldots,M\,, \nonumber \\
G_i^a &=& \delta_i^a \sqrt{\sum_{j=1}^M |\phi_j|^2} \qquad i=1,\ldots,N-1\,.
\end{eqnarray}
When one substitutes Eq.~(\ref{vacuum}) into D-terms one finds that
all D-terms vanish.

In reality the solution of Eq.~(\ref{flat}) is a gauge invariant
surface in field space. In Eq.~(\ref{vacuum}) we have chosen field
values along this flat direction surface, which corresponds to fixing
a particular gauge. Now we have to solve the corresponding gauge field
value from the condition of vanishing matter current, and check whether
it is indeed of pure gauge form. The matter current is given by
\begin{eqnarray}
\label{matter}
J_{\mu}^A &=& i g \left[ \sum_{j=1}^{M} \left( H_j^{\dagger} T^A D_{\mu} H_j
    - (D_{\mu} H_j)^{\dagger} T^A H_j \right) \right. \nonumber \\
& & \left. + \sum_{i=1}^{N-1} \left( G_i^{\dagger} T^A D_{\mu} G_i - (D_{\mu}
    G_i)^{\dagger} T^A G_i \right) \right]
\end{eqnarray}
where the covariant derivative is $D_{\mu} = \partial_{\mu} - ig T^B
A_{\mu}^B$, $g$ is the gauge coupling constant, $A_{\mu}^B$ the
gauge fields, and $T^A$ are the gauge group generators in the defining
representation, which are normalized as; $\rm{Tr} ( T^A T^B ) =
\frac{1}{2} \delta^{AB}$.

Substituting Eq.~(\ref{vacuum}) into the matter current
Eq.~(\ref{matter}), we obtain
\begin{equation}
J_{\mu}^A = ig T^A_{NN} \sum_{j=1}^M \left( \phi_j^* \partial_{\mu} \phi_j -
  \phi_j \partial_{\mu} \phi_j^* \right) + g^2 A_{\mu}^A \sum_{j=1}^M
  |\phi_j|^2 = 0\,.
\end{equation}
Now it is a simple matter to solve the gauge field from this. In order to
check whether it is a pure gauge, we must calculate the field strength
tensor
\begin{equation}
F_{\mu\nu}^A = \partial_{\mu} A_{\nu}^A - \partial_{\nu} A_{\mu}^A + g f^{ABC}
A_{\mu}^B A_{\nu}^C\,.
\end{equation}
Since in the background the scalar fields are function of time alone,
we have only $A_0(t)\neq 0$, rest of the gauge field components
vanish. Therefore $F_{00}^A$ is the only non-vanishing component, but
this also vanishes due to anti-symmetry properties, so we have a pure
gauge configuration, $F_{\mu\nu}^A=0$, as required.

We also assume that the superpotential is such that \eq{vacuum} is also
F-flat, although this is not really necessary, since any superpotential
contributions can be taken into account in the potential. 

%%%%%%%%%%%%%%%%%%%%%%%%%%%%%%%%

\section{Dynamics}

For the purpose of illustration, we keep the potential for the flat
direction, $V\sim f(|\Phi|^{n}/M_{p}^{n-4})$, though we will discuss the
dynamics in a very general context. The Lagrangian for the flat
direction is given by
\begin{eqnarray}
\label{lagr1}
{\cal L} = c\sum_{j=1}^{M} |D_{\mu} H_j|^2 + c\sum_{i=1}^{N-1}
|D_{\mu} G_i|^2 - V(\{H_i,G_j\}),
\end{eqnarray}
where $c=1/2$ for the real fields, and $c=1$ for the complex
fields. By inserting \eq{vacuum} into \eq{lagr1} leads to the
Lagrangian~\footnote{Note that we have left out the gauge field
contribution, since it only affects the dynamics transverse to the
radial motion.}
\begin{eqnarray}
\label{lagr}
{\cal L} = \frac{1}{2} \partial_{\mu}\Phi^{\dagger} \left[1+(N-1)P\right]
\partial^{\mu}\Phi - V(\Phi)\,,
\end{eqnarray} 
where $P = \Phi \Phi^{\dagger} / (\Phi^{\dagger}\Phi)$ is the
projection operator. The field configurations of the real and the
complex fields are
\begin{eqnarray}
\label{Phi1}
\Phi = & (\phi_1,\ldots,\phi_M)^T, & \phi_i \in \mathbb{R}\,, \\
\Phi = & (\phi_1,\ldots,\phi_M,\phi_1^*,\ldots,\phi_M^*)^T, & \phi_i \in
\mathbb{C}\,.
\end{eqnarray}

This Lagrangian results into an equation of motion
\begin{eqnarray}
\label{eqm}
\partial_{\mu}\partial^{\mu}\Phi + 3H\dot\Phi + \frac{N-1}{N}
\frac{\Phi}{\Phi^{\dagger}\Phi} |(1-P)\partial_{\mu}\Phi|^2\nonumber \\ +
\left(1-\frac{N-1}{N}P\right) \dd{V}{\Phi^{\dagger}} = 0\,,
\end{eqnarray}
and the Friedmann equation
\begin{eqnarray}
\label{fried}
3M_p^2H^2 & = \rho =
\frac{1}{2}\dot\Phi^{\dagger}\left[1+(N-1)P\right]\dot\Phi + V \nonumber \\ &
+ \frac{1}{2a^2}\partial_i\Phi^{\dagger}\left[1+(N-1)P\right]\partial_i\Phi\,,
\end{eqnarray}
where $H$ is the Hubble parameter and $a$ is the scale factor.

Now let us concentrate on the radial motion during inflation.  Then it
follows that the centrifugal acceleration vanishes in \eq{eqm}, and we
can write down the equations of motion for
$R=\sqrt{\Phi^{\dagger}\Phi}$ alone,
\begin{eqnarray}
\label{eqm2}
\ddot{R} + 3H\dot{R} + \frac{1}{N} V'(R) = 0\,, \\
\label{fried2}
3M_p^2H^2 = \frac{N}{2} \dot{R}^2 + V(R)\,.
\end{eqnarray}
Now the slow-roll approximation requires that; $|\ddot{R}| \ll 3H|\dot{R}|,~
|V'(R)/N|$ and $N\dot{R}^2/2 \\ \ll V(R)$, so the equations simplify
\begin{eqnarray}
\label{eqm3}
\dot{R} &\approx& -\frac{V'(R)}{3NH}\,, \\
H &\approx & \sqrt{\frac{V(R)}{3M_p^2}}\,.
\end{eqnarray}
An equivalent dynamics can be obtained through the effective slow-roll
parameters
\begin{equation}
\label{eff}
\eps_{eff} = \frac{\eps}{N} \ll 1, \qquad |\eta_{eff}| =
\frac{|\eta|}{N} \ll 1\,,
\end{equation}
where $\eps\equiv(M_{p}^2/2)(V^{\prime}(R)/V(R))^2$ and $\eta\equiv
M_{p}^2(V^{\prime\prime}(R)/V(R))$ are the usual slow roll
parameters. Note an interesting point, both $\epsilon_{eff}$ and
$|\eta_{eff}|$ become less than one for a large number of fields {\it
i.e.} when $N\gg 1$. We can also define the number of e-foldings,
${\cal N}$, as
\begin{equation}
\label{efold}
{\cal N} = \log\left(\frac{a(t_e)}{a(t)}\right)= 
\integral{t}{t_e} H\, dt=\frac{N}{\sqrt{2}M_p} \integral{R_e}{R} dR
\frac{1}{\sqrt{\eps}}\,,
\end{equation}
where the end of inflation is defined by, $\eps_{eff}(R_e)=1$, and $R_e$ is
the value of $R$ at the end of inflation. This shows that even if the
potential were not flat enough for the usual slow-roll conditions to hold, the
effective slow-roll parameters are small enough, therefore the required number
of e-foldings can be generated if there are just enough many $G$ type
fields. In order to solve the flatness and the homogeneity problems we
require ${\cal N} \sim 60$ e-foldings of inflation. This can be achieved by
\begin{equation}
\label{efold2}
N \sim 600 \,\left( \frac{{\cal N}}{60} \right) \, \left( \frac{\Delta R}{0.1
    M_{p}} \right)^{-1} \, \left( \frac{\eps}{2} \right)^{1/2}\,,
\end{equation}
where $\Delta R\equiv R_{i}-R_{e}$. It is evident that we require large 
number of fields in order to realize inflation at vevs lower than 
$M_{p}$.

There are couple of points to be mentioned. First of all it is a good
news that we can drive inflation with gauge invariant flat directions
at vevs smaller than the cut-off scale, this is just one simple
example we were looking for, however, the cost we have to pay is the
large $N$ group. Note here that $N$ is not the dimension of the flat
directions, but it is the number of colours in $SU(N),~SO(N)$ gauge
theories. Further note that the matter content, $N-1+M$, in our case
surpasses the number of colours $N$.

Our solution is still far from realistic. Nevertheless we take an
important message from this analysis; perhaps it is extremely
difficult to seek inflaton sector within a realistic,
phenomenologically interesting, gauge group such as grand unified
groups, e.g. $SU(5),~SO(10)$, etc.

%%%%%%%%%%%%%%%%%%%%%%%%%%%%%%%%%%%%%%%%%%%%%%%%%%%%%%%%%%%%%%%%%%

\section{Density Perturbations}

For the sake of completeness we briefly discuss the density
perturbation arising from the multi-fields.  It is well known how to
calculate the spectrum of the curvature perturbations in the case 
multiple fields, and for a non-flat field metric \cite{SS}
\begin{equation}
\label{spectrum1}
{\cal P}_{\cal R}(k) = \left( \frac{H}{2\pi} \right)^2 (G^{-1})_{ij} \,
\dd{{\cal N}}{\phi_i} \, \dd{{\cal N}}{\phi_j}\,,
\end{equation}
where $G$ is the field metric and the summation goes over field
components running through both $\phi_i$ and $\phi_i^*$ in the case of
complex scalar fields. The isocurvature perturbations have been
ignored in the above analysis. The inverse field metric is
given by
\begin{equation}
\label{inv}
G^{-1} = 1 - \frac{N-1}{N}\,P\,.
\end{equation}
Now assuming that the field trajectory is radial, the number of
e-folds, ${\cal N}$, depends only on the radial motion. From
Eqs.~(\ref{efold},\ref{spectrum1}), the curvature perturbation
spectrum is given by
\begin{equation}
\label{spect}
{\cal P}_{\cal R}(k) = \frac{1}{N}\,\left(\frac{H}{\dot{R}}\right)^2
\left(\frac{H}{2\pi} \right)^2\,=\frac{1}{24\pi^2M_p^4} \frac{V}{\eps_{eff}}\,.
\end{equation}
Note that this last expression contains only the radial mode and it is
the same as the usual formula for a single field case. This reiterates
a point that the dynamics of multi-fields do not alter the spectrum of
density perturbations, see~\cite{Liddle:1998jc}.

Of particular interest is the spectral index $n$. This is given in our case by
\cite{SS}
\begin{equation}
\label{Spectral}
n -1 = -6\epsilon_{eff}+2\eta_{eff}=-\frac{6\epsilon-2\eta}{N}\,.
\end{equation}
This result shows that for larger $N$, the spectral index is even closer
to the scale invariant. 

During inflation there will be fluctuations along the transverse
direction, with an amplitude $\sim H/2\pi$. There are $M-1$ such modes
along which the perturbations will give rise to isocurvature type
fluctuations, whose analysis goes beyond the scope of the present paper.

%%%%%%%%%%%%%%%%%%%%%%%%%%%%%%%%%%%%%%%%%%%%%%%%%%%%%%%%%%%%%%%%%%

\section{Conclusion}

In this letter we addressed the core issue of the inflaton sector; can
we really find a workable model of inflation where the inflaton is
{\it not} a gauge singlet? The answer is positive. Inflation can be
driven by the gauge invariant multi-flat directions. Inflaton in our
case is borne out of a gauge invariant quantity under $SU(N)$ and/or
$SO(N)$ type SUSY gauge theories. We are also able to show that
inflationary scale is sub-Planckian, however it requires a large $N$
of order $600$. This is not a very encouraging news, because as it
appears from our analysis we require a large number of independent
flat directions, which can be obtained only by increasing the number
of colours, $N$, and the matter content.

%%%%%%%%%%%%%%%%%%%%%%%%%%%%%%%%%%%%%%%%%%%%%%%%%%%%%%%%%%%%%%%%%%%
\subsection*{Acknowledgments}
We thank K. Enqvist for a discussion and comments on the manuscript. A.J. is
partly supported by Magnus Ehrnrooth foundation and A.M. is a CITA-National
fellow.

\vskip20pt
               
%%%%%%%%%%%%%%%%%%%%%%%%%%%%%%%%%%%%%%%%%%%%%%%%%%%%%%%%%%%%%%%%%%%

\end{document}